\documentclass[conference]{IEEEtran}
\usepackage{graphicx}
\usepackage{psfrag}
\usepackage{subfigure}
\usepackage{url}
\usepackage{amsmath}
\usepackage{amsfonts,amssymb,amsmath}
\usepackage{array}

\begin{document}

\title{An Improved Scheme for Initial Ranging in OFDMA-based Networks}
\author{\authorblockN{Luca Sanguinetti* and Michele Morelli}
\authorblockA{ \\ Department of Information Engineering \\ University of Pisa
\\Pisa, Italy \\luca.sanguinetti, michele.morelli@iet.unipi.it}\\%
\and \authorblockN{H. Vincent Poor}
\authorblockA{ \\ Department of Electrical Engineering \\\
Princeton University\\
Princeton, NJ USA \\
poor@princeton.edu }} \markboth{Journal of \LaTeX\ Class
Files,~Vol.~1, No.~11,~November~2002}{Shell
\MakeLowercase{\textit{et al.}}: Bare Demo of IEEEtran.cls for
Journals} \maketitle \maketitle
\maketitle \renewcommand\thefootnote{} \footnotetext{*This work was completed
while the author was with Princeton University and it was supported
by the U.S. National Science Foundation under Grants ANI-03-38807 and CNS-06-25637.}

\begin{abstract}
An efficient scheme for initial ranging has recently been proposed by X. Fu
\emph{et al.} in the context of orthogonal frequency-division
multiple-access (OFDMA) networks based on the IEEE 802.16e-2005 standard.
The proposed solution aims at estimating the power levels and timing offsets
of the ranging subscriber stations (RSSs) without taking into account the
effect of possible carrier frequency offsets (CFOs) between the received
signals and the base station local reference. Motivated by the above
problem, in the present work we design a novel ranging scheme for OFDMA in
which the ranging signals are assumed to be misaligned both in time and
frequency. Our goal is to estimate the timing errors and CFOs of each active
RSS. Specifically, CFO estimation is accomplished by resorting to
subspace-based methods while a least-squares approach is employed for timing
recovery. Computer simulations are used to assess the effectiveness of the
proposed solution and to make comparisons with existing alternatives.
\end{abstract}

%


\section{Introduction}

The main impairment of an orthogonal frequency-division multiple-access
(OFDMA) network is represented by its remarkable sensitivity to timing
errors and carrier frequency offsets (CFOs) between the uplink signals and
the base station (BS) local references. For this reason, the IEEE
802.16e-2005 standard for OFDMA-based wireless metropolitan area networks
(WMANs) specifies a synchronization procedure called Initial Ranging (IR)
where subscriber stations that intend to establish a link with the BS can
use some dedicated subcarriers to transmit their specific ranging codes \cite%
{IEEE2006}. Once the BS has revealed the presence of ranging subscriber
stations \ (RSSs), it has to estimate some fundamental parameters including
timing errors, CFOs and power levels.



Two prominent schemes for initial synchronization and power control in OFDMA
were proposed in \cite{Krinock2001} and \cite{Minn2004}. In these works, a
long pseudo-noise sequence is transmitted by each RSS over the available
ranging subcarriers. Timing recovery is then accomplished on the basis of
suitable correlations computed in the frequency- and time-domain,
respectively. The main drawback of these methods is their sensitivity to
multipath distortion, which destroys orthogonality among the employed codes
and gives rise to multiple access interference (MAI). Better results are
obtained in \cite{Lee2005} by using a set of generalized chirp-like (GCL)
sequences, which echibits increased robustness against the channel
selectivity. A different approach to managing the IR process has recently
been proposed in \cite{Minn07}. Here, the pilot streams transmitted by RSSs
are spread in the time-domain over adjacent OFDM blocks using orthogonal
codes. In this way, signals of different RSSs can be easily separated at the
BS as they remain orthogonal after propagating through the channel. Timing
information is eventually acquired in an iterative fashion by exploiting the
autocorrelation properties of the received samples induced by the use of the
cyclic prefix (CP). Unfortunately, this scheme is derived under the
assumption of perfect frequency alignment between the received signals and
the BS local reference. Actually, the occurrence of residual CFOs results
into a loss of orthogonality among ranging codes and may lead to severe
degradations of the system performance in terms of mis-detection probability and
estimation accuracy.

In the present work we propose a novel ranging scheme for OFDMA systems that
is robust to time and frequency misalignments. The goal is to estimate
timing errors and CFOs of all active RSSs. The number of active codes is
found by resorting to the minimum description length (MDL) principle \cite%
{Wax85} while the multiple signal classification (MUSIC) algorithm \cite%
{Schmidt79} is employed to detect which codes are actually active and to
determine their corresponding CFOs. Timing estimation is eventually achieved
through least-squares (LS) methods. Although the proposed solution allows
one to estimate the timing errors of each RSS in a decoupled fashion, it may
involve huge computational burden in applications characterized by large
propagation delays. For this reason, we also present an alternative scheme
derived from ad hoc-reasoning which results into substantial computational
saving. It is worth noting that timing synchronization in OFDMA\ uplink
transmissions has received little attention so far. A well-established way
to handle timing errors is to design the CP length large enough to include
both the channel delay spread and the two-way propagation delay between the
BS and the user station \cite{Morelli2004}. This leads to a
quasi-synchronous system in which timing errors can be viewed as part of the
channel impulse response (CIR) and are compensated for by the channel
equalizer. Unfortunately, this approach poses an upper limit to the maximum
tolerable propagation delay or, equivalently, to the maximum distance
between the BS and the subscriber stations \cite{Pun2007}. For this reason,
its application to a scenario with large cells (as envisioned in next
broadband wireless networks) is hardly viable. In the latter case, accurate
knowledge of the timing errors is required in order to align the uplink
signals to the BS time scale.


\section{System description and signal model}

\subsection{System description}

The investigated OFDMA network employs $N$ subcarriers with frequency
spacing $\Delta f$ and indices in the set $\mathcal{J}=\{0,1,\ldots ,N-1\}$.
Following \cite{Minn07}, we denote $R$ the number of subchannels reserved
for the IR process. Each subchannel is divided into $Q$ subbands uniformly
spaced over the signal bandwidth at a distance $(N/Q)\Delta f$ from each
other. A given subband is composed of a set of $V$ adjacent subcarriers. The
subcarrier indices within the $q$th subband $(q=0,1,\ldots ,Q-1)$ of the $r$%
th ranging subchannel $(r=0,1,\ldots ,R-1)$ are collected into a set $%
\mathcal{J}_{q}^{(r)}=\{i_{q,\nu }^{(r)}~;\nu =0,1,\ldots ,V-1\}$ with
entries
\begin{equation}
i_{q,\nu }^{(r)}=\frac{qN}{Q}+\frac{rN}{QR}+\nu .  \label{1}
\end{equation}%
The $r$th subchannel is thus composed of subcarriers with indices taken from
$\mathcal{J}^{(r)}={\cup _{q=0}^{Q-1}\mathcal{J}_{q}^{(r)}}$. Hence, a total
of $N_{R}=QVR$ ranging subcarriers are available in the system with indices
in the set $\mathcal{J}_{R}={\cup _{r=0}^{R-1}\mathcal{J}^{(r)}}$. The
remaining $N-N_{R}$ subcarriers are used for data transmission and are
assigned to data subscriber stations (DSSs) which have already completed
their IR process and are assumed to be perfectly synchronized to the BS time
and frequency scales \cite{Minn07}.

We denote by $M$ the number of consecutive OFDM blocks reserved for IR and
assume that each ranging subchannel can be accessed at most by $M-1$ RSSs.
The latter are separated by means of specific ranging codes selected in a
pseudo-random fashion from a predefined set $\{\mathbf{c}_{1},\mathbf{c}%
_{2},\ldots ,\mathbf{c}_{M-1}\}$, with $\mathbf{c}_{k}=[c_{k}(1),\,c_{k}(2),%
\ldots ,c_{k}(M)]^{T}$ (the superscript $^{T}$ denotes the transpose
operation). As in \cite{Minn07}, we assume that different RSSs employ
different codes. Without loss of generality, in what follows we concentrate
on the $r$th ranging subchannel and denote by $K^{(r)}\leq M-1$ the number
of simultaneously active RSSs. Also, to simplify the notation, the
subchannel index $^{(r)}$ is dropped in all subsequent derivations.

The waveform transmitted by the $k$th RSS ($1\leq k\leq K$) propagates
through a multipath channel characterized by an impulse response $\mathbf{h}%
_{k}=[h_{k}(0),h_{k}(1),\ldots ,h_{k}(L-1)]^{T}$ of length $L$ (in sampling
periods). At the BS, the received samples are not synchronized with the
local references. We denote by $\theta _{k}$ the timing error expressed in
sampling periods while $\varepsilon _{k}$ is the frequency offset normalized
to the subcarrier spacing. As discussed in \cite{Morelli2004}, subscriber
stations that intend to start the ranging process compute initial frequency
and timing estimates on the basis of a downlink control signal broadcast by
the BS. The estimated parameters are then employed by each RSS as
synchronization references for the uplink ranging transmission. This means
that during IR the CFOs are only due to Doppler shifts and/or estimation
errors and, in consequence, they are assumed to lie \textit{within a small
fraction} of the subcarrier spacing.
Furthermore, in order to eliminate interblock interference (IBI), we assume
that during the ranging process the CP length comprises $N_{G}\geq \theta
_{\max }+L$ sampling periods, where $\theta _{\max }$ is the maximum
expected timing error \cite{Pun2007}. This assumption is not restrictive,
since in many standardized OFDM systems the initialization blocks are
usually preceded by long CPs.

\subsection{Signal model}

We denote by $\mathbf{Y}_{m}$ the $QV$-dimensional vector that collects the
DFT outputs corresponding to the considered subchannel during the $m$th OFDM
block. Since the DSSs are assumed to be perfectly synchronized to the BS
references, their signals will not contribute to $\mathbf{Y}_{m}$. In
contrast, the presence of uncompensated CFOs destroys orthogonality among
ranging signals, thereby leading to some interchannel interference (ICI).
However, as the subchannels are well separated in the frequency domain, we
can reasonably neglect interference on $\mathbf{Y}_{m}$ arising from ranging
signals of subchannels other than the considered one. Under this assumption,
we may write

\begin{equation}
\mathbf{Y}_{m}=\sum\limits_{k=1}^{K}{c_{k}(m)e^{j{m\omega _{k}N_{T}}}%
\mathbf{A}(\omega _{k})\mathbf{S}_{k}}(\theta _{k})+\mathbf{n}_{m}  \label{2}
\end{equation}%
where $\omega _{k} = 2\pi\varepsilon_k /N$, $N_{T}=N+N_{G}$ is the duration
of the cyclically extended block and $\mathbf{n}_{m}$ is a Gaussian vector
with zero mean and covariance matrix $\sigma ^{2}\mathbf{I}_{QV}$ (we denote
by $\mathbf{I}_{N}$ the identity matrix of order $N$). Also, we have defined

\begin{equation}
\mathbf{A}(\omega _{k})=\mathbf{F}\mathbf{V}(\omega _{k})\mathbf{F}^{H}
\label{3}
\end{equation}%
where $\mathbf{V}(\omega _{k})$ accounts for the CFOs and is given by

\begin{equation}
\mathbf{V}(\omega _{k})=\mathrm{diag}\left\{ {e^{j{n}\omega
_{k}};n=0,1,\ldots ,N-1}\right\}  \label{4}
\end{equation}%
while $\mathbf{F}=\left[ {\mathbf{F}_{0}^{H},\mathbf{F}_{1}^{H},\ldots ,%
\mathbf{F}_{Q-1}^{H}}\right] ^{H}$ (the superscript $^{H}$ denotes the
Hermitian transposition) with $\mathbf{F}_{q}$ ($q=0,1,\ldots ,Q-1)$
denoting a $V\times N$ matrix with entries

\begin{equation}
\left[ {\mathbf{F}}_{q}\right] _{\nu ,n}=\frac{1}{\sqrt{N}}e^{-j{2\pi n}%
i_{q,\nu }/N}\quad 0\leq \nu \leq V-1,0\leq n\leq N-1.  \label{5}
\end{equation}%
Vector $\mathbf{S}_{k}(\theta _{k})$ in (\ref{2}) can be partitioned as $%
\mathbf{S}_{k}(\theta _{k}) = \left[ \mathbf{S}_{k}^{T}(\theta
_{k},i_{1}),\mathbf{S}_{k}^{T}(\theta _{k},i_{2}),\ldots ,%
\mathbf{S}_{k}^{T}(\theta _{k},i_{Q-1})\right] ^{T}$, where $\mathbf{S}%
_{k}(\theta _{k},i_{q})$ is a $V$-dimensional vector with elements

\begin{equation}
S_{k}(\theta _{k},i_{q,\nu })=e^{-j{2\pi \theta _{k}}i_{q,\nu
}/N}H_{k}(i_{q,\nu }),\quad 0\leq \nu \leq V-1  \label{6}
\end{equation}%
while $H_{k}(i_{q,\nu })$ denotes the channel frequency response over the $%
i_{q,\nu }$th subcarrier and is given by

\begin{equation}
H_{k}(i_{q,\nu })=\sum_{\ell
=0}^{L-1}h_{k}(\ell )e^{-j{2\pi \ell}i_{q,\nu }/N}.  \label{7}
\end{equation}%
From (\ref{6}) we see that $\theta _{k}$ simply appears as a phase shift
across the DFT outputs. The reason is that the CP length is larger than the
maximal expected propagation delay, thereby making the ranging signals
quasi-synchronous.

In the following sections we show how vectors $\{\mathbf{Y}%
_{m}\}_{m=0}^{M-1} $ can be exploited to compute frequency and timing
estimates for all active ranging codes.

\section{Estimation of the CFOs}

To simplify the derivation, we assume that the CFOs are adequately smaller
than the subcarrier spacing, i.e., $\left\vert {\omega _{k}}\right\vert \ll
1 $. In such a case, matrices $\mathbf{A}(\omega _{k})$ in (\ref{3}) can
reasonably be replaced by $\mathbf{I}_{N}$ to obtain \cite{Morelli2004}

\begin{equation}
\mathbf{Y}_{m}\approx \sum\limits_{k=1}^{K}{c_{k}(m)e^{j{m\omega _{k}N_{T}}%
}\mathbf{S}_{k}}(\theta _{k})+\mathbf{n}_{m}.  \label{9}
\end{equation}%
This equation indicates that each CFO results only in a phase shift between
contiguous OFDM blocks. Collecting the $i_{q,\nu }$th DFT output of all $M$
ranging blocks into a vector $\mathbf{Y}(i_{q,\nu })=[Y_{0}(i_{q,\nu
}),Y_{1}(i_{q,\nu }),\ldots ,Y_{M-1}(i_{q,\nu })]^{T}$, we may write

\begin{equation}
\mathbf{Y}(i_{q,\nu })=\sum_{k=1}^{K}S_{k}(\theta _{k},i_{q,\nu })\mathbf{%
\Gamma }(\omega _{k})\mathbf{c}_{k}+\mathbf{n}(i_{q,\nu })  \label{10}
\end{equation}%
where $\mathbf{n}(i_{q,\nu })$ is Gaussian distributed with zero-mean and
covariance matrix $\sigma ^{2}\mathbf{I}_{M}$, while $\mathbf{\Gamma }%
(\omega _{k})=\text{diag}\{e^{j{m\omega _{k}N_{T}}}~;m=0,1,\ldots ,M-1\}$
is a diagonal matrix that accounts for the phase shifts induced by $\omega
_{k}$.

Inspection of (\ref{10}) reveals that, apart from thermal noise, vector $%
\mathbf{Y}(i_{q,\nu })$ is a linear combination of the frequency-rotated
codes $\{\mathbf{\Gamma }(\omega _{k})\mathbf{c}_{k}\}$. This means that the
signal space is spanned by the $K$ vectors $\{\mathbf{\Gamma }(\omega _{k})%
\mathbf{c}_{k}\}$ that correspond to the active RSSs \cite{StoicaBook97}.
Then, if we temporarily assume that the number $K$ of active codes is known
at the receiver, an estimate of $\omega _{k}$ ($k=1,2,\ldots ,K$) can be
obtained by resorting to the MUSIC algorithm \cite{Schmidt79}. To see how
this comes about, we use the observations $\{\mathbf{Y}(i_{q,\nu })\}$ to
obtain the following sample correlation matrix

\begin{equation}
\hat{\mathbf{R}}_{Y}=\frac{1}{QV}\sum_{v=0}^{V-1}\sum_{q=0}^{Q-1}\mathbf{Y}%
(i_{q,\nu })\mathbf{Y}^{H}(i_{q,\nu }).  \label{14}
\end{equation}%
Next, based on the forward-backward (FB) approach \cite{StoicaBook97}, we
compute

\begin{equation}
\widetilde{\mathbf{R}}_{Y}=\frac{1}{2}(\hat{\mathbf{R}}_{Y}+\mathbf{J}\hat{%
\mathbf{R}}_{Y}^{T}\mathbf{J})  \label{15}
\end{equation}%
where $\mathbf{J}$ is the exchange matrix with 1's on the anti-diagonal and
0's elsewhere. We denote by $\lambda _{1}\geq \lambda _{2}\geq \cdots \geq
\lambda _{M}$ the eigenvalues of $\widetilde{\mathbf{R}}_{Y}$ arranged in
non-increasing order, and by $\{\mathbf{s}_{1},\mathbf{s}_{2},\ldots ,%
\mathbf{s}_{M}\}$ the corresponding eigenvectors. The MUSIC algorithm relies
on the fact that the eigenvectors associated with the $M-K$ smallest
eigenvalues are an estimated basis of the noise subspace and, accordingly,
they are approximately orthogonal to all vectors in the signal space \cite%
{Schmidt79}. Hence, an estimate of $\omega _{k}$ is obtained by minimizing
the projection of $\mathbf{\Gamma }(\widetilde{\omega })\mathbf{c}_{k}$ onto
the noise subspace, i.e.,

\begin{equation}
\hat{\omega }_{k}=\arg \underset{\widetilde{\omega }}{\max }\left\{ \Psi
_{k}(\widetilde{\omega })\right\},  \label{16}
\end{equation}%
with

\begin{equation}
\Psi _{k}(\widetilde{\omega })=\frac{1}{\sum_{m=K+1}^{M}\left\vert \mathbf{c}%
_{k}^{H}\mathbf{\Gamma }^{H}(\widetilde{\omega })\mathbf{s}_{m}\right\vert
^{2}}.  \label{17}
\end{equation}%
It is worth observing that CFO recovery must be accomplished for any active
RSS. However, since the BS has no prior knowledge as to which codes have
been transmitted in the considered subchannel, it must evaluate the
quantities $\{\hat{\omega}_{{1}},\hat{\omega}_{{2}},\ldots ,\hat{\omega}_{{%
M-1}}\}$ for the complete set $\{\mathbf{c}_{1},\mathbf{c}_{2},\ldots ,%
\mathbf{c}_{M-1}\}$. At this stage the problem arises of identifying which
codes are actually active. The identification algorithm looks for the $K$
largest values in the set $\{\Psi _{k}(\hat{\omega}_{k})\}_{k=1}^{M-1}$, say
$\{\Psi _{u_{k}}(\hat{\omega}_{u_{k}})\}_{k=1}^{K}$, and declare as \textit{%
active} the corresponding codes $\{\mathbf{c}_{u_{k}}\}_{k=1}^{K}$. The CFO
estimates are eventually found as $\hat{\boldsymbol{\omega }}_{u}=\mathbf{[}%
\hat{\omega}_{u_{1}},\hat{\omega}_{u_{2}},\ldots ,\hat{\omega}_{u_{K}}%
\mathbf{]}^{T}$.

At this stage we are left with the problem of estimating the parameter $K$
to be used in (\ref{17}). For this purpose, we adopt the\ MDL approach and
obtain \cite{Wax85}

\begin{equation}
\hat{K}=\arg \underset{\tilde{K}}{\min }\left\{ \mathcal{F(}\tilde{K}%
)\right\}  \label{18}
\end{equation}%
where $\mathcal{F(}\tilde{K})$ is the following metric

\begin{equation}
\mathcal{F(}\tilde{K})=\frac{1}{2}\tilde{K}(2M-\tilde{K})\ln (QV) -QV(M-%
\tilde{K})\ln \rho (\tilde{K})  \label{19}
\end{equation}%
with $\rho (\tilde{K})$ denoting the ratio between the geometric and
arithmetic mean of $\{\lambda _{\tilde{K}+1},\lambda _{\tilde{K}+2},\ldots
,\lambda _{M}\}$. Finally, replacing $K$ by $\hat{K}$ in (\ref{17}) leads to
the proposed MUSIC-based frequency estimator (MFE) while the described
identification algorithm is called the MUSIC-based code detector (MCD)

\section{Estimation of the timing delays}

After code detection and CFO recovery, the BS must acquire information about
the timing delays of all ranging signals. This problem is
now addressed by resorting to LS methods. In doing so we still assume that
the number of active codes has been correctly estimated so that $\hat{K}=K$.
Also, to simplify the notation, the indices $\{u_k\}_{k=1}^{K}$ of the
detected codes are relabeled following the map $u_k\longrightarrow k $ for $%
k=1,2,\ldots ,K$.

We begin by reformulating (\ref{10}) in a more compact form. For this
purpose, we collect the CFOs and timing errors in two $K$-dimensional
vectors ${\boldsymbol{\omega }=[}\omega _{1},\omega _{2},\ldots ,\omega _{K}%
\mathbf{]}^{T}$ and $\boldsymbol{\theta }=[\theta _{1},\theta _{2},\ldots
,\theta _{K}]^{T}$. Then, after defining the matrix $\mathbf{C}({\boldsymbol{%
\omega }})=\left[ \mathbf{\Gamma }(\omega _{{1}})\mathbf{c}_{{1}}~~\mathbf{%
\Gamma }(\omega _{{2}})\mathbf{c}_{{2}}~\cdots ~\mathbf{\Gamma }(\omega _{{K}%
})\mathbf{c}_{{K}}\right] $ and the vector $\mathbf{S}(\boldsymbol{\theta }%
,i_{q,\nu })=[S_{{1}}(\theta _{1},i_{q,\nu }),S_{{2}}(\theta _{2},i_{q,\nu
}),\ldots ,S_{{K}}(\theta _{K},i_{q,\nu })]^{T}$, we may rewrite (\ref{10})
in the equivalent form

\begin{equation}
\mathbf{Y}(i_{q,\nu })=\mathbf{C}(\boldsymbol{\omega })\mathbf{S}(%
\boldsymbol{\theta },i_{q,\nu })+\mathbf{n}(i_{q,\nu }).  \label{13}
\end{equation}%
Omitting for simplicity the functional dependence of $\mathbf{S}(\boldsymbol{%
\theta },i_{q,\nu })$ on $\boldsymbol{\theta }$ and assuming $\hat{%
\boldsymbol{\omega }}\approx \boldsymbol{\omega }$, from (\ref{13}) the
maximum likelihood estimate of $\mathbf{S}(i_{q,\nu })$ is found to be

\begin{equation}
\hat{\mathbf{S}}(i_{q,\nu })=[\mathbf{C}^{H}(\hat{\boldsymbol{\omega }})%
\mathbf{C}(\hat{\boldsymbol{\omega }})]^{-1}\mathbf{C}^{H}(\hat{\boldsymbol{%
\omega }})\mathbf{Y}(i_{q,\nu }).  \label{18}
\end{equation}%
Substituting (\ref{13}) into (\ref{18}) yields

\begin{equation}
\hat{\mathbf{S}}(i_{q,\nu })=\mathbf{S}(i_{q,\nu })+\boldsymbol{\xi }%
(i_{q,\nu })  \label{19}
\end{equation}%
where $\boldsymbol{\xi }(i_{q,\nu })$ is a zero-mean disturbance term. From (%
\ref{6}) and (\ref{7}) it follows that

\begin{equation}  \label{20}
\hat{S}_{{k}}(i_{q,\nu })=e^{-j\frac{{2\pi \theta_{k}}}{N}i_{q,\nu
}}\sum_{\ell =0}^{L-1}h_{k}(\ell )e^{-j\frac{{2\pi n}}{N} i_{q,\nu }}+\xi _{{%
k}}(i_{q,\nu }).
\end{equation}%
On denoting $\hat{\mathbf{S}}_{{k}}(\nu)=\left[ \hat{S}_{{k}}(i_{0,\nu }),%
\hat{S}_{{k}}(i_{1,\nu }),\ldots ,\hat{S}_{{k}}(i_{Q-1,\nu })\right] ^{T}, $
and $\boldsymbol{\Phi }(\theta _{{k}},\nu)=\mathrm{diag}\{e^{-j\frac{{2\pi
\theta _{{k}}}}{N}i_{q,\nu }}~;~q=0,1,\ldots ,Q-1\}$, we may rewrite (\ref%
{20}) as follows

\begin{equation}
\hat{\mathbf{S}}_{{k}}(\nu)=\boldsymbol{\Phi }(\theta _{{k}},\nu)\mathbf{F}%
(\nu)\mathbf{h}_{{k}}+\boldsymbol{\xi }_{{k}}(\nu)  \label{21}
\end{equation}%
where $\boldsymbol{\xi }_{{k}}(\nu)=[\xi _{{k}}(i_{0,\nu }),\xi _{{k}%
}(i_{1,\nu }),\ldots ,\xi _{{k}}(i_{Q-1,\nu })]^{T}$ while $\mathbf{F}(\nu)$
is a matrix of dimension $Q\times L$ with entries $\lbrack \mathbf{F}%
(\nu)]_{q,\ell }= e^{-j\frac{{2\pi \ell }}{N}i_{q,\nu }}$ for $0\leq q\leq
Q-1$ and $0\leq \ell \leq L-1$ 

Equation (\ref{21}) indicates that, apart from the disturbance term $%
\boldsymbol{\xi }_{{k}}(\nu )$, $\hat{\mathbf{S}}_{{k}}(\nu )$ is only
contributed by the ${k}$th RSS, meaning that ranging signals have been
successfully decoupled at the BS. We may thus exploit vectors $\{\hat{%
\mathbf{S}}_{{k}}(\nu )~;~\nu =0,1,\ldots ,V-1\}$ to get LS estimates of $%
(\theta _{{k}},\mathbf{h}_{{k}})$ separately for each RSS. This amounts to
minimizing the following objective function with respect to $(\tilde{\theta}%
_{{k}},\tilde{\mathbf{h}}_{{k}})$

\begin{equation}
\Lambda _{{k}}(\tilde{\theta}_{{k}},\tilde{\mathbf{h}}_{{k}%
})=\sum_{\nu=0}^{V-1}\left\Vert \hat{\mathbf{S}}_{{k}}(\nu) - \boldsymbol{%
\Phi }(\tilde{\theta}_{{k}},\nu)\mathbf{F}(\nu)\tilde{\mathbf{h}}_{{k}%
}\right\Vert ^{2}.  \label{23}
\end{equation}%
For a fixed $\tilde{\theta}_{{k}}$, the minimum of $\Lambda _{{k}}(\tilde{%
\theta}_{{k}},\tilde{\mathbf{h}}_{{k}})$ is achieved at

\begin{equation}
\hat{\mathbf{h}}_{{k}}=\frac{1}{QV}\sum_{\nu=0}^{V-1}\mathbf{F}^{H}(\nu)%
\boldsymbol{\Phi }^{H}(\tilde{\theta}_{{k}},\nu)\hat{\mathbf{S}}_{{k}}(\nu)
\label{24}
\end{equation}%
where we have used the identity $\mathbf{F}^{H}(\nu)\mathbf{F}(\nu)=Q \cdot
\mathbf{I}_{L}$. Then, substituting (\ref{24}) into (\ref{23}) and
minimizing with respect to $\tilde{\theta}_{k}$ yields the timing estimate
in the form

\begin{equation}
\hat{\theta}_{k}=\arg \underset{0\leq \tilde{\theta}_{k}\leq \theta _{\max }}%
{\max }\left\{ \Upsilon (\tilde{\theta}_{k})\right\}  \label{25}
\end{equation}%
where $\Upsilon (\tilde{\theta}_{k})$ is given by

\begin{equation}
\Upsilon (\tilde{\theta}_{k})=\sum_{\ell =\tilde{\theta}_{k}}^{\tilde{\theta}%
_{k}+L-1}\left\vert \sum_{\nu=0}^{V-1}\hat{s}_{k}(\nu,\ell )e^{j2\pi \ell
\nu/N}\right\vert ^{2}  \label{n32/4}
\end{equation}%
and we have denoted by $\hat{s}_{k}(\nu,\ell )$ the $Q$-point IDFT of the
sequence $\{\hat{S}_{k}(i_{q,\nu });0\leq q\leq Q-1\}$. In the sequel $\hat{%
\theta}_{k}$ is termed the LS-based timing estimator (LS-TE).

Once $\hat{\theta}_{k}$ has been computed from (\ref{25}), it is used in (%
\ref{24}) to estimate the CIR of the $k$th RSS as

\begin{equation}
\hat{\mathbf{h}}_{k}=\frac{1}{QV}\sum_{\nu=0}^{V-1}\mathbf{F}^{H}(\nu)%
\boldsymbol{\Phi }^{H}(\hat{\theta}_{k},\nu)\hat{\mathbf{S}}_{k}(\nu).
\label{24.1}
\end{equation}%
It is worth noting that for $V=1$ the timing metric (\ref{n32/4}) reduces to

\begin{equation}
\left. \Upsilon (\tilde{\theta}_{k})\right\vert _{V=1}=\sum_{\ell =\tilde{%
\theta}_{k}}^{\tilde{\theta}_{k}+L-1}\left\vert \hat{s}_{k}(0,\ell
)\right\vert ^{2}  \label{n33}
\end{equation}%
and becomes periodic in $\tilde{\theta}_{k}$ with period $Q$. In such a
case, the estimate $\hat{\theta}_{k}$ is affected by an ambiguity of
multiples of $Q$. This ambiguity does not represent a serious problem as
long as $Q$ can be chosen to be greater than $\theta _{\max }$.
Unfortunately, in some applications this may not be the case. For example,
in \cite{Minn07} we have $Q=8$ while $\theta _{\max }=102$.

\subsection{Reduced-complexity timing estimation}

Although separating the RSS signals at the BS considerably reduces the
system complexity, evaluating $\Upsilon (\tilde{\theta}_{k})$ for $\tilde{%
\theta}_{k}=0,1,\ldots ,\theta _{\max }$ may still be computationally
demanding, especially in applications where $\theta _{\max }$ is large. For
this reason, we now develop an ad-hoc reduced complexity timing estimator
(RC-TE).

We begin by decomposing the timing error $\theta _{k}$ into a fractional
part $\beta _{k}$, less than $Q$, plus an integer part which is multiple of $%
Q$, i.e.,

\begin{equation}
\theta _{k}=\beta _{k}+p_{k}Q  \label{n34}
\end{equation}%
where $\beta _{k}\in \{0,1,\ldots ,Q-1\}$ while $p_{k}$ is an integer
parameter taken from $\{0,1,\ldots ,P-1\}$ with $P=\left\lfloor \theta
_{\max }/Q\right\rfloor $. Omitting the details, it is possible to rewrite $%
\Upsilon (\tilde{\theta}_{k})$ as

\begin{equation}
\Upsilon (\tilde{\theta }_{k})=\Upsilon _{1}(\tilde{\beta }_{k})+\Upsilon
_{2}(\tilde{\beta }_{k},\tilde{p}_{k})  \label{n35}
\end{equation}%
where
\begin{equation}
\Upsilon _{1}(\tilde{\beta }_{k})=\sum_{\ell =\tilde{\beta }_{k}}^{\tilde{%
\beta }_{k}+L-1}\sum_{\nu=0}^{V-1}\left\vert \hat{s}_{k}(\nu,\ell
)\right\vert ^{2}  \label{n36}
\end{equation}
while $\Upsilon _{2}(\tilde{\beta }_{k},\tilde{p}_{k})$ is shown in (\ref%
{n36bis}) at the top of the next page.
\begin{figure*}[htp]
\begin{equation}
\Upsilon _{2}(\tilde{\beta }_{k},\tilde{p}_{k})=2\Re e \left\{\sum_{\ell =%
\tilde{\beta }_{k}}^{\tilde{\beta }_{k}+L-1}\sum_{\nu=0}^{V-2}\sum_{n=1}^{V-%
\nu-1}\hat{s}_{k}(\nu,\ell )\hat{s}_{k}^{\ast }(\nu+n,\ell )e^{-j2\pi n(\ell
+\tilde{p}_{k}Q)/N}\right\}  \label{n36bis}
\end{equation}%
\end{figure*}

The RC-TE is a suboptimal scheme which, starting from (\ref{n35}), estimates
$\beta _{k}$ and $p_{k}$ in a decoupled fashion. More precisely, an estimate
of $\beta _{k}$ is first obtained looking for the maximum of $\Upsilon _{1}(%
\tilde{\beta}_{k})$, i.e.,

\begin{equation}
\hat{\beta}_{k}=\arg \underset{0\leq \tilde{\beta}_{k}\leq Q-1}{\max }%
\left\{ \Upsilon _{1}(\tilde{\beta}_{k})\right\} .  \label{n40}
\end{equation}%
Next, replacing $\beta _{k}$ with $\hat{\beta}_{k}$ in the right-hand-side
of (\ref{n35}) and maximizing with respect to $\tilde{p}_{k}$, provides an
estimate of $p_{k}$ in the form

\begin{equation}
\hat{p}_{k}=\arg \underset{0\leq \tilde{p}_{k}\leq P-1}{\max }\left\{
\Upsilon _{2}(\hat{\beta }_{k},\tilde{p}_{k})\right\}.  \label{n41}
\end{equation}
A further reduction of complexity is possible when $V=2$. Actually, in this
case it can be shown that maximizing $\Upsilon _{2}(\hat{\beta}_{k},\tilde{p}%
_{k})$ is equivalent to maximizing $\cos (\varphi _{k}-2\pi \tilde{p}%
_{k}Q/N) $, where

\begin{equation}
\varphi _{k}=\arg \left\{ \sum_{\ell =\hat{\beta}_{k}}^{\hat{\beta}_{k}+L-1}%
\hat{s}_{k}(0,\ell )\hat{s}_{k}^{ \ast }(1,\ell )e^{-j2\pi \ell /N}\right\} .
\label{n44}
\end{equation}%
The estimate of $p_{k}$ is thus obtained in closed-form as

\begin{equation}
\hat{p}_{k}=\frac{N\varphi _{k}}{2\pi Q}.  \label{n45}
\end{equation}

\section{Simulation results}

\subsection{System parameters}

The simulated system is inspired by \cite{Minn07}. The total number of
subcarriers is $N=1024$ while the number of ranging subchannels is $R=8$.
Each subchannel is composed by $Q=16$ subbands uniformly spaced at a
distance $N/Q=64$. Any subband comprises $V=2$ adjacent subcarriers while
the ranging time-slot includes $M=4$ OFDM blocks. The ranging codes are
taken from a Fourier set of length $4$ and are randomly selected by the RSSs
at each simulation run (expect for the sequence $\left[ 1,1,1,1\right] ^{T}$%
). The discrete-time CIRs have $L=12$ channel coefficients. The latter are
modeled as circularly symmetric and independent Gaussian random variables
with zero means (Rayleigh fading) and exponential power delay profiles,
i.e., $E\{\left\vert h_{k}(\ell )\right\vert ^{2}\}=\sigma _{h}^{2}\cdot
\exp (-\ell /12)$ with $\ell =0,1,\ldots ,11$ and $\sigma _{h}^{2}$ chosen
such that $E\{\left\Vert \mathbf{h}_{k}\right\Vert ^{2}\}=1$. Channels of
different users are assumed to be statistically independent. They are
generated at each new simulation run and kept fixed over an entire
time-slot. The normalized CFOs are uniformly distributed within the interval
$[-\Omega ,\Omega ]$ and vary at each simulation run. We consider a cell
radius of 10 km, corresponding to a maximum transmission delay $\theta
_{\max }=204$. A CP of length $N_{G}=256$ is chosen to avoid IBI.

\begin{figure}[t]
\begin{center}
\includegraphics[width=.45\textwidth]{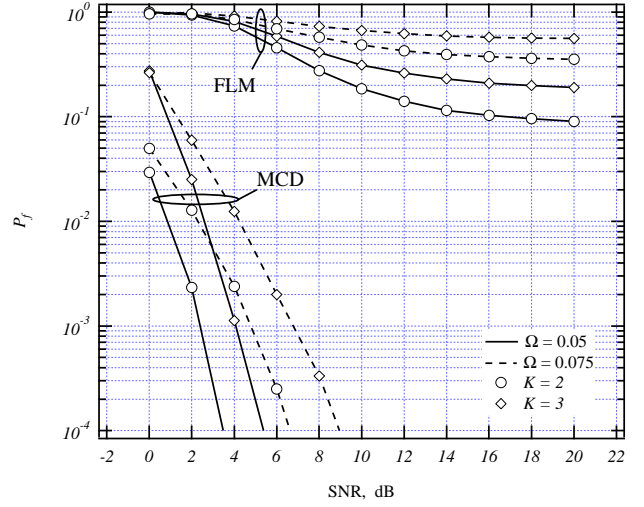}
\end{center}
\caption{$P_{f}$ vs. SNR for $K$ = 2 or 3 when $\Omega$ is 0.05 or 0.075.}
\label{picture1}
\end{figure}


\subsection{Performance evaluation}

We begin by investigating the performance of MCD in terms of probability of
making an incorrect detection, say $P_{f}$. This parameter is illustrated in
Fig. 1 as a function of SNR = $1/ \sigma^2$ under different operating
conditions. The number of active RSSs varies from 2 to 3 while the maximal
frequency offset is either $\Omega =0.05$ or $0.075$. Comparisons are made
with the ranging scheme discussed by Fu, Li and Minn (FLM) in \cite{Minn07},
where the $k$th ranging code is declared active provided that the quantity

\begin{equation}
{\mathcal{Z}}_{k}=\frac{1}{M^2}\sum\limits_{q=0}^{Q-1}{\sum\limits_{%
\nu=0}^{V-1}{\left\vert {\mathbf{c}_{k}^{H}\mathbf{Y}(i_{q,\nu})}\right\vert
^{2}}}
\end{equation}%
exceeds a suitable threshold $\eta $ which is proportional to the estimated
noise power $\hat{\sigma}^{2}$. The results of Fig. 1 indicates that
the proposed scheme performs remarkably better than FLM because of its
intrinsic robustness against CFOs. As expected, the system performance
deteriorates for large values of $K$ and $\Omega $. The reason is that
increasing $K$ reduces the dimensionality of the noise subspace, which
degrades the accuracy of the MUSIC estimator. Furthermore, large CFO values
result into significant ICI which is not accounted for in the signal model (%
\ref{9}), where $\mathbf{A}(\omega _{k})$ has been replaced by $\mathbf{I}%
_{N}$.

\begin{figure}[t]
\begin{center}
\includegraphics[width=.45\textwidth]{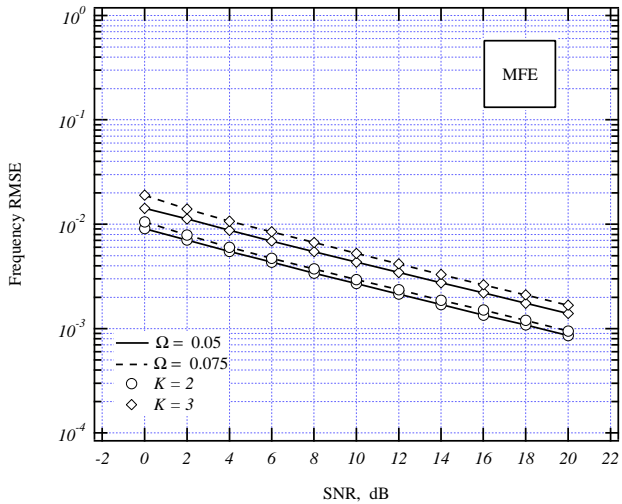}
\end{center}
\caption{Accuracy of the frequency estimates vs. SNR for $K$ = 2 or 3 when $\Omega$ is 0.05 or 0.075. }
\label{picture1}
\end{figure}
\begin{figure}[t]
\begin{center}
\includegraphics[width=.45\textwidth]{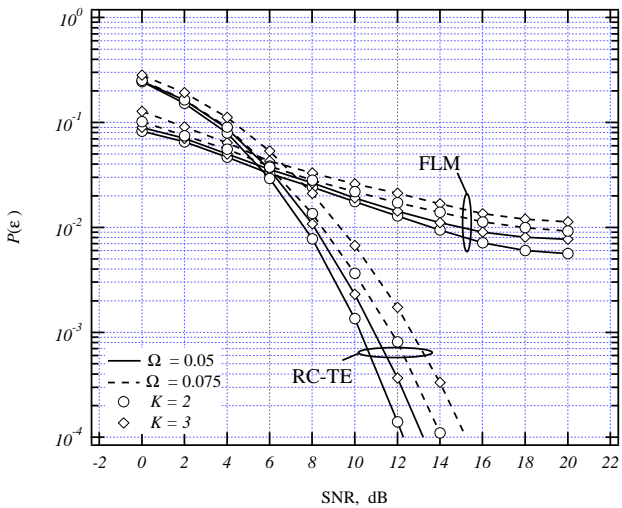}
\end{center}
\caption{$P(\protect\epsilon)$ vs. SNR for $K$ = 2 or 3 when $\Omega$ is
0.05 or 0.075.}
\label{picture1}
\end{figure}

Fig. 2 illustrates the root mean-square-error (RMSE) of the frequency
estimates obtained with MFE vs. SNR. Again, we see that the system
performance deteriorates when $K$ and $\Omega $ are relatively large.
Nevertheless, the accuracy of MFE is satisfactory under all investigated
conditions.

The performance of the timing estimators is measured in terms of probability
of making a timing error, say $P(\epsilon )$, as defined in \cite%
{Morelli2004}. More precisely, an error event is declared to occur whenever
the estimate $\hat{\theta}_{k}$ gives rise to IBI during the data section of
the frame. In such a case, the quantity $\hat{\theta}%
_{k}-\theta _{k}+(-N_{G,D}+L)/2$ is larger than zero or smaller than $%
-N_{G,D}+L-1$, where $N_{G,D}$ is the CP length during the data transmission
phase. Fig. 3 illustrates $%
P(\epsilon )$ vs. SNR as obtained with RC-TE and FLM when $N_{G,D}=64$. The operating
conditions are the same of the previous figures. Since the performance of
LS-TE is virtually identical to that of RC-TE, it is not reported in order
not to overcrowd the figure. We see that for SNR values larger than $6$ dB the proposed scheme provides
much better results than FLM.



\section{Conclusions}

We have derived a novel timing and frequency synchronization scheme for
initial ranging in OFDMA-based networks. The proposed solution aims at
detecting which codes are actually being employed and provides timing and
CFO estimates for all active RSSs. CFO estimation is accomplished by
resorting to the MUSIC algorithm while a LS approach is employed for timing
recovery. Compared to the timing synchronization algorithm discussed in \cite%
{Minn07}, the proposed scheme is more robust to frequency misalignments and
exhibits improved accuracy.

\bibliographystyle{IEEEtran}
\bibliography{IEEEabrv,../../../../../../Bibliography/bibnew}

\end{document}